\documentstyle[aps,prb,epsf,floats]{revtex}
\begin{document}

\twocolumn[\hsize\textwidth\columnwidth\hsize\csname@twocolumnfalse%
\endcsname

\draft
\title{The quantum Coulomb glass within the Hartree-Fock approximation }

\author{Frank Epperlein, Michael Schreiber and Thomas Vojta}
\address{Institut f\"ur Physik, Technische Universit\"at, D-09107 Chemnitz, Germany}
\date{version April 8, printed \today}
\maketitle

\begin{abstract}
We study the influence of electron-electron
interactions on the electronic properties of disordered 
materials. In particular, we consider the insulating side of a 
metal-insulator transition where screening breaks down and the 
electron-electron inter\-action remains long-ranged. 
The investigations are based on the quantum Coulomb glass, 
a generalization of the classical Coulomb glass model
of disordered insulators. The quantum Coulomb glass is studied by decoupling the 
Coulomb interaction by means of a 
Hartree-Fock approximation 
and exactly diagonalizing the remaining localization problem.
We investigate the behavior of the Coulomb gap in the density of states
when approaching the
metal-insulator transition and study the influence of the interaction 
on the localization of the electrons. We find that the interaction leads to an enhancement of
localization at the Fermi level.

\end{abstract}

]  


\section{Introduction}

The influence of electron-electron interactions on the electronic properties of disordered 
systems has reattracted a lot of attention recently. Already disorder alone can lead to a 
metal-insulator transition (MIT) by means of spatial localization of the electronic states
at the Fermi energy.
This MIT, called the Anderson transition, has been investigated extensively within the last two
decades.\cite{kramrev93} While the qualitative features of the Anderson transition are
well understood by now, the description remains inconsistent at a quantitative level. In
particular, the critical behavior is not completely understood and the results of several methods
do not agree.

Moreover, today it is generally assumed that the MIT in most experimental systems
cannot be described by a model of non-interacting electrons since the Coulomb interaction
between the electrons plays a crucial role. 
The metallic regime of the disordered interacting electron system is comparatively well
understood, at least qualitatively.\cite{belitzrev94} Altshuler and Aronov showed \cite{aa79}
that the single-particle density of states (DOS) displays a non-analyticity at the Fermi energy
which was called the Coulomb anomaly. Later the perturbative treatment was extended
into the whole metallic phase by means of a field-theoretic renormalization group method
\cite{belitzrev94,finkel83} which permits a qualitative discussion of the MIT 
including the identification of the different universality classes. However, quantitative
results are very difficult to obtain from these methods. This is in particular so since the
$\epsilon$-expansion which has to be used to extrapolate to the physical dimension
$d=3$ is highly singular.\cite{epsexp}

Whereas investigations in the metallic phase can be carried out by means of  established
diagrammatic methods analogous studies of the insulating phase are not possible.
That is because the natural reference system for a perturbation theory, viz. a system having 
disorder and interactions but no overlap between the states at different sites, 
is an interacting system and diagrammatic methods cannot be applied since Wick's theorem
does not hold.
Instead, the insulating limit itself represents a challenging many-body problem.
Almost three decades ago Pollak predicted \cite{pollak70} an 
interaction-induced reduction of the single-particle DOS at the Fermi
energy in disordered insulators. Later Efros and Shklovskii defined \cite{es75}
the prototype model of disordered electronic systems in the insulating limit, the
classical Coulomb glass model. They showed that the zero-temperature single-particle
DOS has power-law gap at the Fermi energy which is called the Coulomb gap.\cite{es75,efros76}
This suggests the question whether Coulomb anomaly and 
Coulomb gap are manifestations of the same physical phenomenon on the
metallic and insulating sides of the MIT, respectively. We will come back to this 
question in Sec. III.
The physics of the classical Coulomb glass model has been investigated 
in much detail by several analytical and numerical methods and its static properties 
are comparatively well understood by now.\cite{cgnum}
In contrast, the nature of the transport mechanism is still 
controversially discussed.\cite{pollak92}

Since experiments deep in the insulating regime are difficult to carry out
most results on disordered insulators have been obtained from samples
not too far away from the MIT.\cite{exp} Here the (single-particle) localization
length is still much larger than the typical distance between two sites
and the description of the electrons in terms of classical point charges becomes
questionable. Attempts to include the overlap between different states
into the Coulomb glass model
have been made earlier \cite{cpa} by mapping the problem
onto a non-interacting model and applying the coherent potential approximation. 
However, in this method neither disorder nor interactions are treated 
completely and different results obtained this way contradict each other. 
Recently, localization in an interacting disordered system was
  investigated \cite{talamantes96} by the numerical
analysis of  the many-body spectrum of small clusters from which the
authors inferred a delocalizing influence of the interactions.

We note, that in addition to these works which deal with the ground state properties
of many-body systems possessing a finite particle density there has been
a very active line of research concerning the behavior of just two interacting
particles in a random environment.\cite{tip} This type of work concentrates
on special highly excited states of the two-electron system which will,
in general, behave differently from the ground state at finite particle density.

In this paper we investigate the physics of the disordered interacting
electron problem (having finite particle density) on the insulating side of 
the MIT.  In order to account for a finite overlap between the states we 
generalize the classical Coulomb glass model to a quantum model
by including transfer matrix elements between different sites. We then study
two main questions: (i) How does the single-particle DOS and, in particular, the 
Coulomb gap depend on the transfer between the sites? (ii) How does the 
Coulomb interaction influence the localization of the electrons? 

The paper is 
organized as follows: In Sec. II we define the quantum Coulomb glass model and
explain our calculational method. In Sec. III we present the results for the 
single-particle DOS and discuss the behavior of the Coulomb gap. 
The localization properties and the resulting phase diagram of the MIT are
considered in Sec. IV and Sec. V is devoted to some discussions and
conclusions.

\section{The quantum Coulomb glass model}

In the insulating limit the overlap between the electronic states at different sites can be 
neglected and the electrons behave like classical point charges. The generic model
for this regime is the classical Coulomb glass model \cite{es75} which consists of
classical point charges in a random potential which interact via Coulomb interactions.
The model is defined on a regular hypercubic lattice with $N=L^d$ ($d$ is the spatial dimensionality) 
sites occupied by $K N$ (spinless) 
electrons ($0\!<\!K\!<\!1$). To ensure charge neutrality
each lattice site carries a compensating positive charge of  $Ke$. The Hamiltonian
of the classical Coulomb glass reads
\begin{equation}
H_{\rm cl} = \sum_i (\varphi_i - \mu) n_i + \frac{1}{2}\sum_{i\not=j}(n_i-K)(n_j-K)U_{ij}~.
\end{equation}
Here $n_i$ is the occupation number of site $i$ and $\mu$ 
is the chemical potential. The Coulomb interaction $U_{ij} = e^2/r_{ij}$
remains long-ranged since screening breaks down in the insulating phase. 
We set the interaction strength of nearest neighbor sites to 1 which fixes the
energy scale.  
The random potential values $\varphi_i$ are chosen 
independently from a box
distribution of width $2 W_0$ and zero mean.

Our goal is  to describe the regime where the overlap between the states
at different sites cannot be neglected but the system is still insulating. 
Therefore, we generalize 
the Coulomb glass model model to a quantum Coulomb glass model
by adding hopping matrix elements between nearest neighbors.
The Hamiltonian of the quantum Coulomb glass is given by
\begin{equation}
H =  -t  \sum_{\langle ij\rangle} (c_i^\dagger c_j + c_j^\dagger c_i) + H_{\rm cl},
\label{eq:Hamiltonian}
\end{equation}
where $c_i^\dagger$ and $c_i$ are the electron creation and annihilation operators
at site $i$, respectively,  and the sum runs over all pairs of nearest neighbor sites. 
In the limit $t \rightarrow 0$ the model (\ref{eq:Hamiltonian}) reduces to the classical
Coulomb glass, for vanishing Coulomb interaction but finite overlap it reduces to the
usual Anderson model of localization.

In order to calculate the electronic properties we decouple the Coulomb interaction
by means of a Hartree-Fock approximation giving
\begin{eqnarray}
H_{\rm HF} =&-&t  \sum_{\langle ij\rangle} (c_i^\dagger c_j + c_j^\dagger c_i) 
+  \sum_i  (\varphi_i -\mu) n_i \nonumber \\ 
 &+& \sum_{i\not=j} n_i ~ U_{ij} \langle n_j -K \rangle  
\label{eq:HF} \\
&-& \sum_{i,j} c_i^\dagger c_j ~ U_{ij} \langle c_j^\dagger c_i \rangle, \nonumber
\end{eqnarray}
where the first two terms contain the single-particle part of the Hamiltonian, the
third is the Hartree energy and the fourth term contains the exchange interaction.
(Note that in (\ref{eq:HF})  several constant terms have been dropped.)
$\langle \ldots \rangle$ represents the expectation value with respect to
the Hartree-Fock ground state which has to be determined self-consistently. 
In this way the quantum Coulomb glass
is reduced to a self-consistent disordered single-particle problem which we 
solve by means of numerically exact diagonalization giving the single-particle 
energies $\varepsilon_\nu$ and states $|\psi_\nu\rangle$.
We note that the Hartree-Fock approximation is exact for both of the limiting
cases mentioned above, viz. the classical Coulomb glass and the Anderson
model of localization.

In this study we investigate three-dimensional quantum Coulomb glass
systems with up to $N=10^3$ 
sites and band fillings $K$ between 1/2 and 15/16. Due to the particle-hole symmetry 
of the Hamiltonian (\ref{eq:Hamiltonian}) this also covers the band fillings between 1/16 and 1/2.
The disorder strength is fixed at $W_0=1$, and
the overlap parameter $t$ varies from zero (classical limit) up to $t=0.5$ which is 
above the MIT.  In order to reduce the statistical error we average the results over
100 different configurations of the random potential $\varphi$.

\section{Single-particle density of states}
\label{sec:III}

For several reasons, the single-particle DOS plays a special role in the
investigation of the quantum Coulomb glass. First, it is the quantity  investigated
best for the classical Coulomb glass where it shows the well-known power-law
Coulomb gap.\cite{es75,efros76} One question we want to address in this section is 
whether the Coulomb gap remains intact in the presence of  a small overlap. This
question is of central importance for the justification of the classical model in experiments
comparatively close to the MIT where the overlap between different impurity states cannot 
be neglected. Second, from field-theoretic studies on the metallic side of the MIT it was 
inferred \cite{belitz94} that the single-particle DOS at the Fermi energy is the order parameter
of the disorder-driven MIT in {\em interacting} systems.  Thus it should remain zero in the whole
insulating phase and start to increase when crossing the MIT point. Third, in the metallic phase
the DOS should display the aforementioned Coulomb anomaly, a square-root non-analyticity
on top of a finite background.

\begin{figure}
\epsfxsize=8cm
\centerline{\epsffile{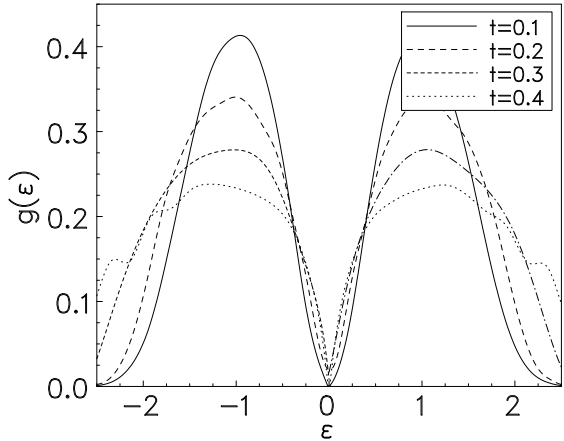}}
\caption{Density of states of the quantum Coulomb glass for
 $W_0\!=\!1$, $K\!=\!0.5$.}
\label{fig:dos05}
\end{figure}

\begin{figure}
\epsfxsize=8cm
\centerline{\epsffile{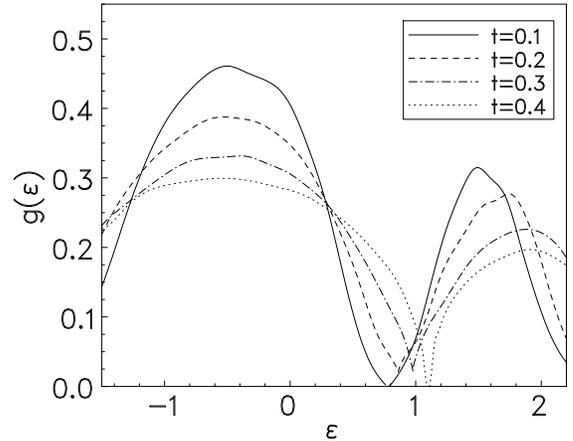}}
\caption{Density of states of the quantum Coulomb glass for
 $W_0\!=\!1$, $K\!=\!0.75$.}
\label{fig:dos025}
\end{figure}

In the context of the Hartree-Fock approximation the single-particle energies 
are simply given by the eigenvalues $\varepsilon_\nu$ 
of the self-consistent Hartree-Fock Hamiltonian (\ref{eq:HF}).
Thus, the single-particle DOS is defined by
\begin{equation}
g(\varepsilon) = \frac 1 N \sum_\nu \delta(\varepsilon - \varepsilon_\nu)~.
\end{equation}
Our numerical results for the single-particle DOS of the quantum Coulomb glass 
are comprised in Figs. \ref{fig:dos05} and \ref{fig:dos025}.
Note that the Hamiltonian is particle-hole symmetric for $K=0.5$
(Fig. \ref{fig:dos05}). Thus the Fermi energy does not depend on $t$. For $K=0.75$, in contrast,
the Fermi energy increases with $t$, and the shift of the gap position in Fig. \ref{fig:dos025} exactly
matches the shift of the Fermi energy. 

In order to address the questions raised at the beginning of this section we study the behavior 
of the DOS close to the Fermi energy.  Let us first discuss our expectations: 
In the insulating phase the electrons are localized and cannot screen the Coulomb interaction.
Consequently,  the Hartree part of the interaction remains long-ranged while the exchange part
which is proportional to the overlap of different states is small and short-ranged. 
In the insulating phase we can therefore apply  a generalization of 
the Efros-Shklovskii argument to discuss the behavior of the leading terms of  the DOS: 
The original argument  \cite{es75,efros76} shows that an empty state $j$ and an 
occupied state $i$ with an energetic distance smaller than $\delta$ must have a spatial distance
larger than $e^2/\delta$ since the change $\Delta=\varepsilon_j - \varepsilon_i -e^2/r_{ij}$ 
of the system energy when moving the electron from $i$ to $j$ must be positive in the many-body 
ground state.
If finite overlaps between different sites are included the electrons become somewhat delocalized 
and therefore the interaction is screened on {\em short} length scales of the order of the localization length. 
In contrast, the long-range part of the interaction remains unchanged. 
Since the DOS close to the Fermi energy is determined by the long-range tail of the interaction we expect 
it to remain unchanged as long as the electrons are localized.
However, the region of validity of the classical result shrinks to zero with increasing delocalization and
vanishes when the electronic states become extended. Thus the Coulomb gap should become
narrower with increasing $t$ and vanish at the MIT. 

On the other hand, with increasing delocalization of the electrons the exchange interaction 
becomes larger and longer-ranged. Since the exchange interaction is responsible  for
the Coulomb anomaly \cite{anomaly} we expect the DOS to show a crossover from
the Coulomb gap behavior to a Coulomb anomaly behavior. 
To be precise, if the system is close to the MIT but still insulating the DOS should
show Coulomb-gap-like behavior in a narrow interval around the Fermi energy
and Coulomb-anomaly-like behavior for energies a bit away from the Fermi level.

\begin{figure}
\epsfxsize=8cm
\centerline{\epsffile{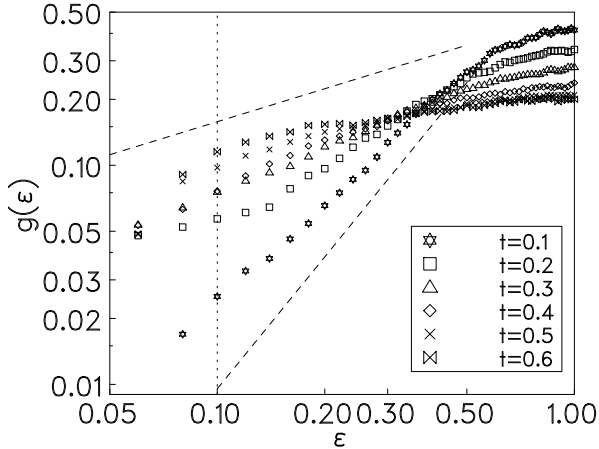}}
\caption{Log--log plot of the density of states of the quantum Coulomb glass for
 $W_0\!=\!1$, $K\!=\!0.5$. The lower dashed line represents the analytical result  \protect\cite{efros76}
 for the DOS in the classical model ($t=0$), the upper dashed line corresponds to a square-root 
 behavior with arbitrary prefactor and
 the dotted line is the reliability limit due to the cut-off
 of the Coulomb interaction, see text.}
\label{fig:doslog}
\end{figure}

In Fig. \ref{fig:doslog} we present a log--log plot of the single-particle DOS in the Coulomb gap region for the 
system with band filling $K=0.5$.
The data presented are compatible with the above expectations. For small $t$ we find a power-law
behavior with an exponent close to 2 as expected for the Coulomb gap in 3D.  With increasing $t$
the exponent becomes smaller and approaches 0.5 as expected for the Coulomb anomaly (if
the constant background is small). We are, however, not able
to explicitly demonstrate the crossover from the Coulomb gap to the Coulomb anomaly
at fixed $t$ as a function of energy . The main reason is that  the investigation of the DOS 
very close to the Fermi energy is hampered by strong finite size effects. The usual problem, viz.
that a finite system always possesses a discrete spectrum, is  made worse by  the long-range character 
of the interaction.
Since the maximum system size considered here is $L=10$ the Coulomb interaction is
effectively cut-off at distances $r_{ij} \sim 10$ which corresponds to $U_{ij} \sim 0.1$. Thus  the results 
for energies smaller than $\varepsilon \sim 0.1$ are not reliable.

\section{Localization properties}

The usual criteria for localization are defined for non-interacting electrons only, and
their generalization to many-body systems is not straightforward. Within the Hartree-Fock
approximation, however, we do obtain effective single-particle states and energies
so that the usual localization criteria can be applied.

\subsection{Participation number}
\label{sec:partnum}

\begin{figure}
\epsfxsize=8cm
\centerline{\epsffile{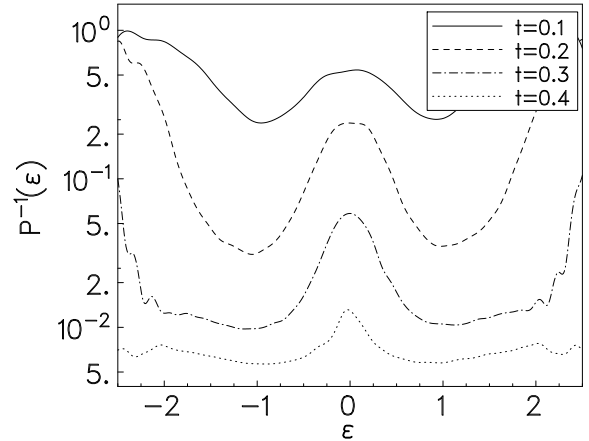}}
\caption{Inverse participation number of the quantum Coulomb glass as a function of energy 
 for  $W_0\!=\!1$, $K\!=\!0.5$.}
\label{fig:part05}
\end{figure}

\begin{figure}
\epsfxsize=8cm
\centerline{\epsffile{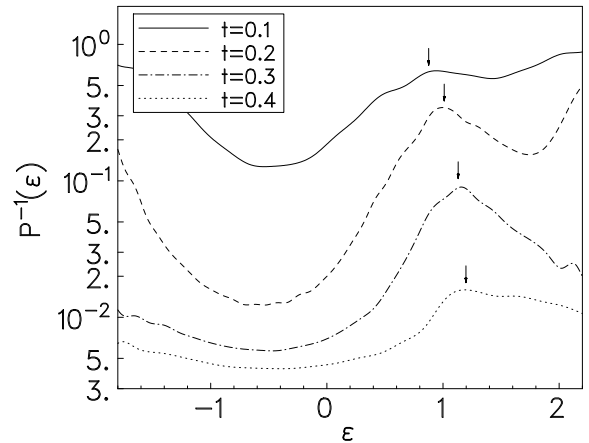}}
\caption{Inverse participation number 
  of the quantum Coulomb glass as a function of energy for
 $W_0\!=\!1$, $K\!=\!0.75$. The arrows mark the positions of the Fermi energy.}
\label{fig:part025}
\end{figure}

One of the simplest measures to study the localization properties is the participation
number $P$ which describes how many sites are effectively occupied by a 
single-particle state $|\psi_\nu\rangle$. Thus the inverse participation number measures the degree of
localization. It is defined as the second moment
of the spatial probability distribution of the state
\begin{equation}
P^{-1}_\nu = \frac{1}{N} \sum_i |\langle \psi_\nu | i \rangle |^4
\end{equation}
where the sum runs over all sites $i$.
In practice it is often averaged over all states with a certain energy $\varepsilon$
\begin{equation}
P^{-1}(\varepsilon) = \frac 1 {g(\varepsilon)}\: \frac 1 N \sum_\nu P^{-1}_\nu\: \delta(\varepsilon-\varepsilon_\nu)~.
\label{eq:partnum}
\end{equation}
In Figs. \ref{fig:part05} and \ref{fig:part025} we show the results for the inverse participation numbers
of systems with $N=10^3$ sites and band filling factors of $K=0.5$ and 0.75, respectively.
The most remarkable feature of these results is the strong enhancement of 
$P^{-1}$ close to the Fermi energy which can be as large as one order of
magnitude (note the logarithmic scale in the figures). This enhancement of $P^{-1}$ 
corresponds to a much stronger localization at the Fermi level compared
to the rest of the band. It is a direct  consequence  of the Coulomb gap 
in the DOS which means a reduction of the number of states that can be hybridized by 
a certain overlap $t$. 
Based on this argument it is also easy to understand how the enhancement
depends on the overlap $t$: For very small $t$  all states remain strongly 
localized so that there is no room for a large enhancement.
The largest enhancement is obtained for moderate values of $t$ which are still
smaller than the width of the Coulomb gap. In this case the states away from the
Fermi level are considerably delocalized while the hybridization at the Fermi energy 
is still hampered. For overlaps $t$ larger than the width of the Coulomb gap
hybridization becomes easier also at the Fermi level and thus the enhancement of 
$P^{-1}$ is diminished. 
Note that in contrast to non-interacting electrons the participation numbers 
depend on the band filling since the electronic states are influenced by
the interaction with the other electrons.

\begin{figure}
\epsfxsize=8cm
\centerline{\epsffile{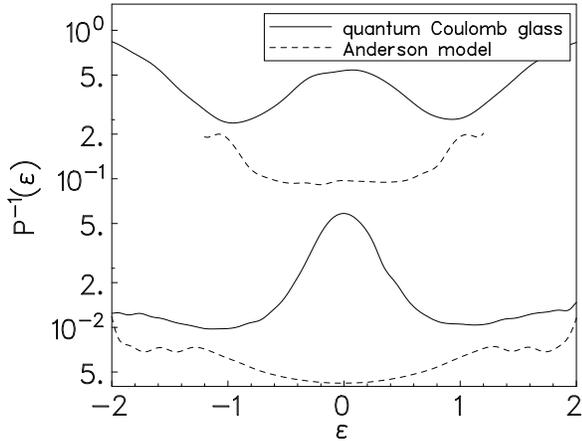}}
\caption{Comparison of the inverse participation number  
 of the quantum Coulomb glass ($K\!=\!0.5$) with that of the
 Anderson model of localization for
 $t=0.1$ (top) and $t=0.3$ (bottom).}
\label{fig:partand}
\end{figure}

A comparison (see Fig. \ref{fig:partand}) of the inverse participation numbers of the
quantum Coulomb glass and of non-interacting electrons shows that in the interacting system
the electrons are more strongly localized all over the band but the 
enhancement is strongest close to the Fermi energy.

Although the inverse participation number is a useful quantity to study 
qualitative features of  localization it is not well suited to quantitatively determine
the MIT and its properties. The reason is that  determining the MIT from the 
participation numbers amounts to detecting changes in the {\em size dependence}
of $P^{-1}$ which is much harder than detecting changes of $P^{-1}$ itself.
($P$ should remain finite for $N\to \infty$ for localized states but scale with $N$
for extended states.)
We therefore use a different method based on the properties of the eigenvalue 
spectrum of the Hamiltonian
which is explained in the following subsection.

\subsection{Level statistics}

The mobility edge, i.e. the energy that separates extended from localized states, 
can be found by using the statistical properties of the energy levels as was done
for the Anderson model of localization.\cite{levshklov,levhof}
In this method the distribution $P$ of nearest-neighbor level spacings
$s$ of the (unfolded) spectrum of eigenvalues $\varepsilon_\nu$ is considered. 
In accordance with the literature we use the notation $P(s)$ for this distribution,
it should not be confused with the participation number $P$ discussed in Sec. \ref{sec:partnum}.
At the MIT the level spacing distribution 
function displays a sharp  transition from the
Poisson ensemble (PE, for the insulating phase) via 
the critical ensemble (at the transition point) to the Gaussian orthogonal ensemble
(GOE, in the metallic phase). For finite system sizes a smooth crossover between the three
ensembles is observed instead of the sharp transition for the infinite system.

Within the Hartree-Fock approximation the quantum Coulomb glass is equivalent
to an Anderson model of localization having an unusual disorder distribution  and
additional disordered transfer elements. As in the Anderson model,  the MIT can 
therefore be determined by the crossover of the level
spacing distribution $P(s)$. 
 
\begin{figure}
\epsfxsize=8cm
\centerline{  \epsffile{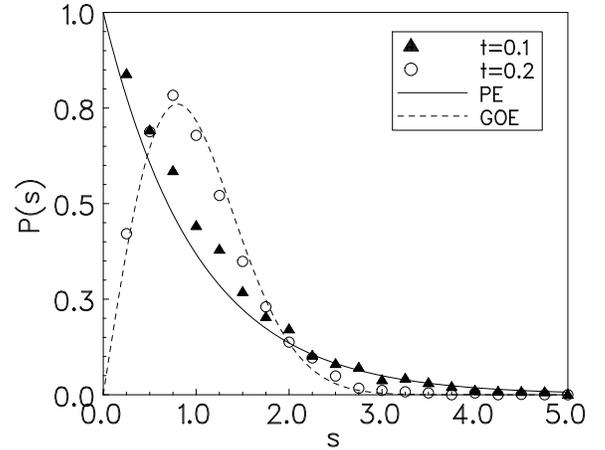}}
 \caption{Level spacing distribution $P(s)$ of the quantum Coulomb glass at the energy
                 $\varepsilon=0.9$ compared to PE and GOE.
               The data are taken from a system with $10^3$ sites, a band filling 
               of $K=0.5$ and a disorder strength $W_0=1$.}
\label{fig:levstat}
\end{figure}

\begin{figure}
\epsfxsize=8cm
\centerline{\epsffile{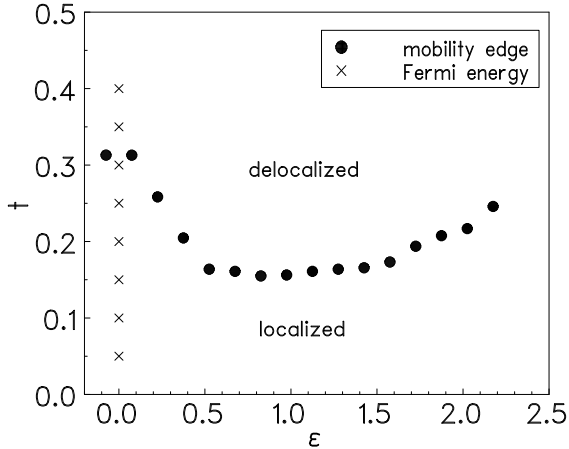}}
\caption{Mobility edge of the quantum Coulomb glass for
 $W_0\!=\!1$, $K\!=\!0.5$.}
\label{fig:k05mit}
\end{figure}

\begin{figure}
\epsfxsize=8cm
\centerline{\epsffile{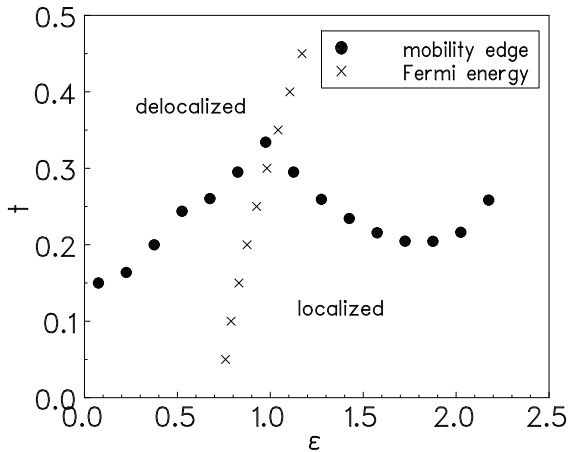}}
\caption{Mobility edge of the quantum Coulomb glass for
 $W_0\!=\!1$, $K\!=\!0.75$.}
\label{fig:k025mit}
\end{figure}

In Fig. \ref{fig:levstat} we show examples for the distribution. For $t=0.1$ the spectrum is close to
PE (the states are localized as will be shown later) while for $t=0.2$ the spectrum is close to 
GOE (the states are extended).
In order to  determine the location of the MIT in parameter space 
the crossover from PE to GOE has to be described 
quantitatively. Following Ref. \cite{levhof} we fit the numerically obtained distributions
$P(s)$ to the phenomenological formula \cite{fitformel}
\begin{equation}
P_{phe}(s) = As^\beta(1+C \beta s)^{f(\beta)} \exp \left[ -\frac {\pi^2} {16} \beta s^2
    - \frac \pi 4 (2-\beta) s  \right ]
\end{equation}
with $f(\beta) = 2^\beta (1-\beta/2)/\beta-0.16874$.  This formula interpolates smoothly between 
PE and GOE. It contains only a single 
free parameter since the first two moments of the level spacing distribution $P(s)$ are
normalized:
\begin{equation}
\int ds \: P(s) = \int ds\: s \: P(s) =1 ~.
\end{equation}
We then study the dependence of the fit parameters $A$, $C$ and $\beta$ on the 
single-particle energy $\varepsilon$ and overlap strength $t$.
The parameter $\beta$ shows a particular strong dependence close to the
mobility edge. From Ref.\cite{levhof} it is known that the critical ensemble 
corresponds to $\beta \approx 0.875$ which we use as a criterion to
determine the transition point. The resulting dependence of the mobility
edge on energy and overlap is presented in Figs. \ref{fig:k05mit} and
\ref{fig:k025mit}.
Close to the Fermi energy the mobility edge is shifted to larger overlaps 
(or, equivalently, smaller disorder), so the location of the mobility edge also
reflects the enhancement of localization at the Fermi energy.

\subsection{Metal--insulator transition}

\begin{figure}
  \epsfxsize=8cm
\centerline{  \epsffile{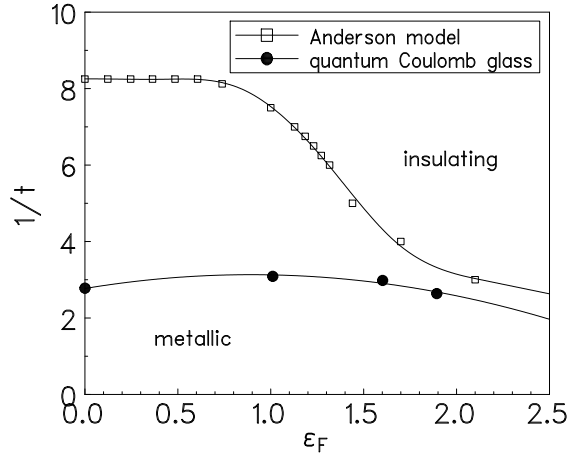}}
  \caption{Phase diagram of the MIT of the quantum Coulomb glass
   and the Anderson model for $W_0=1$. The lines are guides to the eyes only.} 
  \label{fig:pd}
\end{figure}

In a system of non-interacting electrons the states and energy levels 
do not depend on the filling of the band. Changing the filling factor 
simply leads to a shift of the Fermi energy within the otherwise
unchanged band. When the Fermi energy crosses the (fixed) mobility edge
the system undergoes a  MIT. 
In a system of interacting electrons, however, the mobility edge
changes with filling factor $K$.
Therefore, separate calculations have to be done for different
filling factors to determine the phase diagram. The MIT
occurs when the states {\em at the Fermi energy} delocalize (or localize).
This means that Figs. \ref{fig:k05mit} and \ref{fig:k025mit} yield only one
data point each for the phase boundary.
We have carried out the corresponding calculations  for filling
factors $K=7/8$ and 15/16, too. The resulting phase
diagram of the MIT is displayed in Fig. \ref{fig:pd} and compared
to the analogous phase diagram for the Anderson model of
localization.\cite{bulka}
We find that the phase boundary of the quantum Coulomb
glass is shifted to significantly larger values of the overlap $t$ compared to non-interacting electrons
if the Fermi energy is well within the band.
As discussed above this is a direct consequence of the Coulomb gap in the single-particle 
DOS. For band fillings close to $K=1$ the critical $t$ of the quantum Coulomb glass remains 
almost unchanged since the form of the Coulomb gap does not depend on its position in the 
band whereas the critical $t$ of the Anderson model is reduced because
the DOS of the Anderson model decreases near the band edges.

\section{Conclusions}

To summarize, we have investigated the combined influence of disorder and interactions
on the properties of electronic systems on the insulating side of the MIT.  Our work is
based on  the quantum Coulomb glass model.  We have decoupled the interaction by means of 
the Hartree-Fock approximation and numerically diagonalized the remaining disordered
single-particle problem. The resulting single-particle DOS shows a Coulomb gap in the
whole insulating phase which becomes narrower when approaching the MIT.
The reduced DOS at the Fermi energy leads to an enhancement of localization  compared 
to the rest of the band and also compared to non-interacting electrons.

In this concluding section we will discuss some aspects of the results that 
have not yet been covered.
First, we want to discuss the justification of the Hartree-Fock approximation.
On a qualitative level, there are several possible influences of the Coulomb interaction
on Anderson localization with competing effects. On the one hand, the Coulomb interaction 
leads to a reduction of the density of states at the Fermi energy which enhances
localization. This process is contained in the Hartree-Fock approximation as is discussed
in Sec. \ref{sec:III} and as
we have demonstrated in this paper.  On the other hand one may argue that any
interaction leads to transitions between the states of the non-interacting system 
thus giving the electrons additional hopping possibilities and reducing localization.
This second point is not well described within the Hartree-Fock approximation. 
We have therefore started to compare the results of this paper to that of 
exact diagonalizations of small lattices. Preliminary results \cite{exakt} show that
the large enhancement of  localization at the Fermi level is also found by the exact diagonalizations
while the average degree of localization in the band is overestimated by the Hartree-Fock 
approximation
in some parameter regions. Further studies along these lines are in progress.
We also note that, as is well known,  the Hartree-Fock approximation of the 3D homogeneous interacting
electron system produces an artificial soft gap at the Fermi energy since screening is not treated
properly.  Although this artificial gap is much narrower ($g \sim 1/\ln|\varepsilon-\varepsilon_F|$) 
than the Coulomb anomaly and thus difficult to observe the results for the DOS on the 
{\em metallic}  side of the MIT  may be influenced and more sophisticated investigations
will have to be carried out.

Second, we want to comment on the relation between Coulomb gap in the insulating phase
and Coulomb anomaly in the metallic phase. It has been suggested that both are different
manifestations of the same physical phenomenon.  However, the Coulomb gap is a result
of the Hartree part of the interaction and its existence is
tied to the long-range nature of the Coulomb interaction. In contrast, the Coulomb anomaly 
is produced by the exchange interaction and arises independently of the range even for point-like interactions. 
Therefore a system with
a short-range model interaction will display a Coulomb anomaly but not a Coulomb gap.
Further work is necessary to clarify how the Coulomb anomaly changes to the Coulomb gap 
at the MIT (in the case of long-range interactions) or  how it vanishes on the insulating side for short-range
interactions.

This work was supported in part by the DFG under Grant Nos. SFB 393, Schr231/13-1 and 
Vo659/1-1.


\end{document}